# High Data Rate Laser Communications for the Black Hole Explorer

Jade Wang[a], Bryan Bilyeu[a], Don Boroson[a], Dave Caplan[a], Kat Riesing[a], Bryan Robinson[a], Curt Schieler[a], Michael D. Johnson[b], Lindy Blackburn[b], Kari Haworth[b], Janice Houston[b], Sara Issaoun[b], Daniel Palumbo[b], Elliot Richards[b], Ranjani Srinivasan[b], Jonathan Weintroub[b], Dan Marrone[c]

[a]MIT Lincoln Laboratory, [b]Center for Astrophysics | Harvard & Smithsonian, [c]University of Arizona

## ABSTRACT

The Black Hole Explorer (BHEX) is a mission concept that can dramatically improve state-of-the-art astronomical very long baseline interferometry (VLBI) imaging resolution by extending baseline distances to space. To support these scientific goals, a high data rate downlink is required from space to ground. Laser communications is a promising option for realizing these high data rate, long-distance space-to-ground downlinks with smaller space/ground apertures. Here, we present a scalable laser communications downlink design and current lasercom mission results.

## 1. INTRODUCTION

Laser communications is poised to enable a new era of scientific missions by providing significantly more data return in small size, weight and power packages. The investments and technological maturation over the past several decades have now come to fruition. Laser communications systems have now been reduced to operational use and today carry 100's Gbps across a proliferated low-earth-orbit (LEO) backbone for terrestrial commercial services. In addition, the European Space Agency (ESA) EDRS system operates a relay in geosynchronous orbit (GEO) which performs 1000's of laser communications links from LEO to GEO in a month [1-2]. Over the past decade, laser communications has been demonstrated to deliver reliable, error-free data transmission from space to ground from LEO, GEO, cis-lunar, and deep space [3-7].

With the significant cost and long timelines for developing space science missions, maximizing the data return of those missions is of critical importance. In addition, some science missions are only feasible with the data rates provided by laser communications. The Black Hole Explorer (BHEX) mission concept will take advantage of this maturation of laser communications capability to dramatically improve state-of-the-art astronomical Very Long Baseline Interferometry (VLBI) of black holes with only a modest size radio-frequency antenna [8-10].

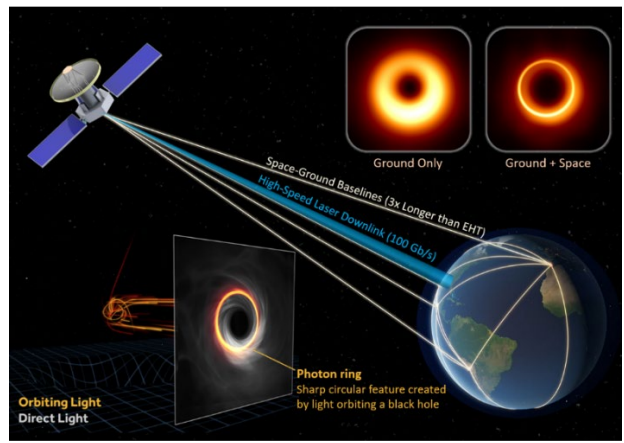

Figure 1. Overview of the Black Hole Explorer mission concept (reproduced from [10]).

DISTRIBUTION STATEMENT A. Approved for public release. Distribution is unlimited.This material is based upon work supported by the MIT - Unspecified under Air Force Contract No. FA8702-15-D-0001. Any opinions, findings, conclusions or recommendations expressed in this material are those of the author(s) and do not necessarily reflect the views of the MIT - Unspecified. © 2024 Massachusetts Institute of Technology.Delivered to the U.S. Government with Unlimited Rights, as defined in DFARS Part 252.227-7013 or 7014 (Feb 2014). Notwithstanding any copyright notice, U.S. Government rights in this work are defined by DFARS 252.227-7013 or DFARS 252.227-7014 as detailed above. Use of this work other than as specifically authorized by the U.S. Government may violate any copyrights that exist in this work.

This mission will answer fundamental open questions on black hole physics such as

- What are the spins of supermassive black holes?
- What powers the feedback from supermassive black holes?
- How do black holes grow?

The ability to uniquely image the black hole with sufficient resolution to measure the photon ring relies on achieving sufficient sensitivity and long baselines greater than the diameter of the earth [8,10]. Sampling and downlinking the observational spectrum at 64 Gbps enables a successful mission with only a ~3m class antenna, thus enabling ground-breaking scientific discoveries in a small, low cost mission such as the NASA Small Explorer class missions.

To enable this vision, in 2021 we developed a scalable architecture based on high TRL technologies [11]. A design was developed for a GEO-based satellite capable of downlinking data from 16 Gbps up to 256 Gbps – well beyond what is required. To support the required error-free transmission of data captured at 64 Gbps with a cost-effective, low size weight and power system leveraging high TRL technologies, the laser communications system will support a downlink operating at 100 Gbps. This design relied on two major developments that showed the maturity of laser communications technologies for supporting BHEX: the high data rate communications module, based on the Terabyte Infrared Delivery (TBIRD) mission, and the large aperture optical terminal suitable for higher orbits, based on the Modular, Agile, Scalable Optical Terminal (MAScOT) developed for two NASA missions. In addition, since the original study was undertaken, the lasercom space industry has begun to mature and additional options for optical terminals are now beginning to be offered on the commercial market. In section 2, we give an update on the results of the TBIRD mission and the MAScOT terminal. In section 3, we give an update on the BHEX lasercom downlink requirements and outline the implications on the laser communications system design.

## 2. TBIRD: 200-GBPS CLASS DATA DOWNLINK

TBIRD (Terabyte Infrared Delivery) is a cubesat mission demonstrating the potential of laser communications systems to deliver large volumes of science data from space to ground [4,12]. The short wavelengths (high carrier frequencies) of the optical spectrum results in lower diffraction losses and more directed beams in free space, allowing for smaller apertures to achieve the same gains as radio-frequency systems. In addition, optical wavelengths provide a larger spectrum and more data bandwidth. Optical wavelengths, however, are more susceptible to atmospheric impairments and laser communication systems between space and ground use a variety of techniques to mitigate these impairments such as beam diversity, interleaving-and-coding, and adaptive optics systems. TBIRD demonstrated high data rates (200 Gbps), small size, weight and power (a 3L, 3kg cubesat payload), and a novel atmospheric mitigation technique that can be used with commercial low-cost high bandwidth terrestrial transceivers.

In addition, this technique demonstrates a new "buffer and burst" architecture which enables large data volume delivery without the need for substantial ground infrastructure [13]. By buffering data on board with a 2-TB buffer, a sensor generating megabytes of data can collect and store the data on board over multiple orbits. Then, when the appropriate optical ground station is available, TBIRD can downlink the data in a burst, emptying its entire buffer of science data directly to a ground station that can be co-located with the science analysis center. Analysis of the results from the cubesat demonstrator indicated that 6 TB could potentially be downlinked in a single pass to a small ground terminal with only a 25-cm aperture [12].

The TBIRD mission launched in 2022 and the payload is still in operation as of July 2023. Recent results showed the capability of downlinking up to 4.8 TB of data in a single ~5-minute pass (Fig 2). This demonstration shows the capability of laser communications in revolutionizing science missions and bringing about a new era of scientific discovery. The BHEX mission will leverage this capability to deliver substantial new scientific discoveries with a modest-sized space mission from a Medium Earth Orbit (MEO) of 20,000 km. As this is substantially longer range than the TBIRD LEO ~500 km orbit, the BHEX lasercom system will need a larger space aperture.

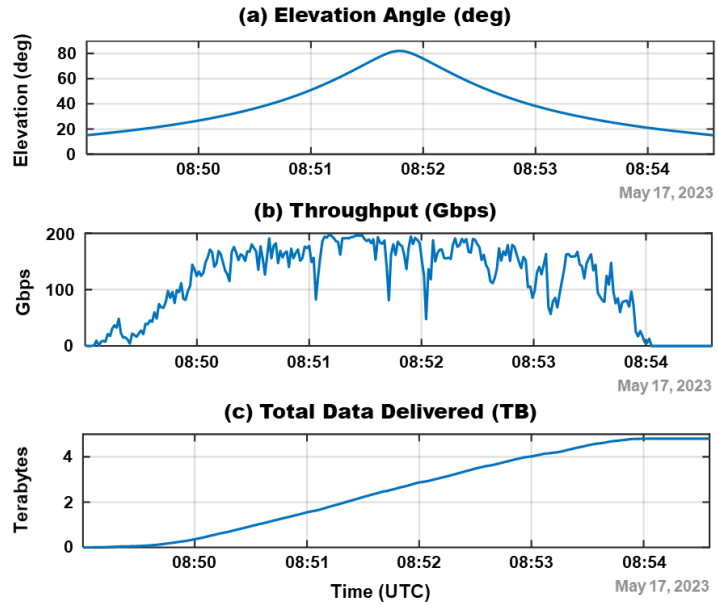

Figure 2. Results from a single TBIRD low-earth-orbit pass over the ground station reproduced from [12]. (a) shows the elevation angle over the duration of the pass. (b) shows the throughput across the pass. The throughput changes as the atmospheric and ground station conditions change. (c) shows the total cumulative data delivered.

## 3. MAScOT: 10-CM OPTICAL TERMINAL

Many of the considerations of a laser communications are similar to those of a radio frequency link. One key aspect for both is the power/aperture/range design trade. To close a free-space communications link from space to ground, the space aperture, ground aperture, and transmit power must be sufficient to overcome the range loss. A space to ground downlink can use a larger ground aperture to offset a smaller space aperture, or lower transmit power. In addition to technical trades, cost is often a major consideration for the design of scientific satellite missions. Such costs are driven by size, mass, power, complexity as well as by technology and design maturation. New development of terminals can be risky and expensive. In addition, with optical telescopes, as the size increases, the beam size generally decreases and the pointing accuracy must also increase. Thus, there is a trade in terms of maturity of telescope design, pointing accuracy, and ground terminal complexity.

One optical terminal with substantial heritage and sufficient aperture for the BHEX mission data downlink requirements is the Modular, Agile, Scalable Optical Terminal (MAScOT) [14]. The MAScOT is a versatile optical telescope with pointing, acquisition and tracking (PAT) controller designed to support a wide variety of missions from low-earth-orbit to cislunar with a single versatile design. The MAScOT provides a hemispherical field-of-regard and a modular design allowing for industry build and ease of integration. Two versions of this terminal were built. The first version was built to support the ILLUMA-T mission, which provides an optical communications link from the International Space Station to the NASA Laser Communications Relay Demonstration (LCRD) currently in operation in GEO [5]. As of publication, this terminal is currently demonstration successful operations between the ISS, LCRD and optical ground stations.

This same terminal design is also used as part of the Artemis II mission and will demonstrate the versatility of the design to support laser communication systems across a wide range of orbits [15]. This terminal has successfully gone through environmental testing and is currently integrated on the Artemis spacecraft, which is due to launch in 2025.

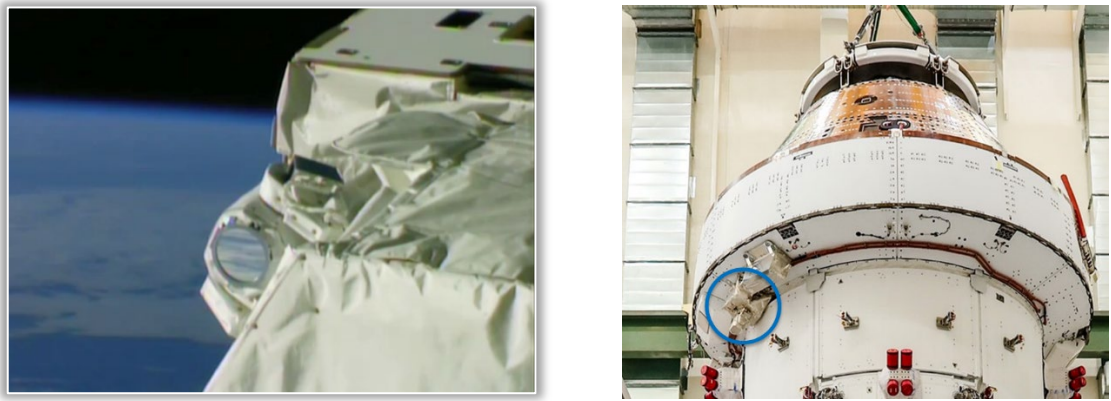

Figure 3. Left: A photograph of the MAScOT terminal on the International Space Station. Right: A second MAScOT terminal integrated on the Artemis II capsule, planned to launch in 2025.

In addition to the MAScOT, other terminals of sufficient aperture size have been demonstrated on orbit and additional terminals are beginning to come on the market and are expected to launch within the next few years [1-3, 5-7, 14]. One such example of optical communications terminals with substantial heritage, though largely in the 1um regime, are the terminals developed by the European Space Agency for the European Data Relay Satellite system (EDRS) [1]. These terminals have been in successful operation for many years in both LEO and GEO orbits, delivering optical communications from space to ground. Over the several years, as the next generation of optical communications systems begin to be fielded and additional commercial offerings mature and become available, we expect the options for BHEX laser communications optical telescopes to increase.

## 4. LASER COMMUNICATIONS DOWNLINK SYSTEM DESIGN

In this section we give an update to the BHEX laser communications system design. A system diagram is shown in Figure 4. The satellite lasercom modem module will ingest data from the BHEX sensor at the required ~64 Gbps. This modem module will need a small buffer to support the TBIRD-based atmospheric-mitigation scheme. We calculate that this buffer will be less than 16 GB to provide error-free communications, and can be smaller if we allow additional errors as VLBI correlation is tolerant to high errors. A high-power optical amplifier will provide the signal power required to close the link range from MEO (~20,000 km). An optical telescope and pointing, acquisition and tracking controller will maintain the optical communications link. A low-rate (~kbps) uplink signal and pointing beacon will also be needed. On the ground, at each optical ground station, a large aperture telescope will collect the signal and feed it to a tracking system and an adaptive optics system (required for apertures in the 30+cm regime). The optical signal will be coupled into a single mode fiber and the optical communications modem hardware demodulate the signal and provide the appropriate sensor data for analysis. The data will either be stored on site or relayed back to the science analysis team for correlation.

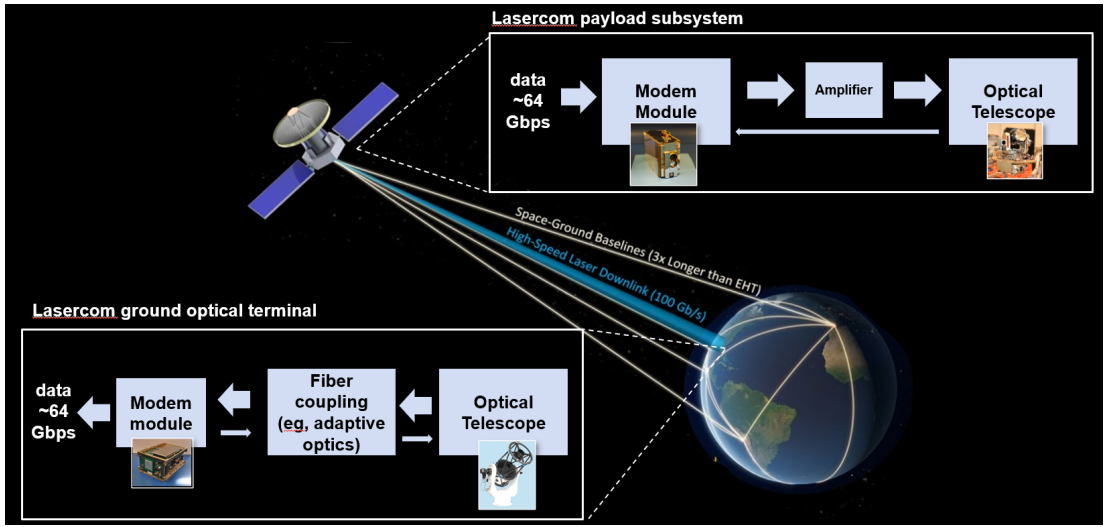

Figure 4. System architecture of the lasercom downlink for the Black Hole Explorer (BHEX) mission.

With the substantial relaxation of requirements from the initial architecture, we can now broaden the trade space and allow for our system design to take better advantage of recent commercial and government developments in maturing laser communications systems. Table 1 presents a few design options for the laser communications system, focusing on a range of achievable space apertures, ground apertures, and high power space-qualified amplifiers.

Table 1. Potential design points within the BHEX lasercom downlink trade space. Smaller space apertures will require higher transmit amplifier power and larger ground apertures. Larger space apertures can enable smaller, cheaper ground terminals potentially avoiding the need for adaptive optics.

| Transmit Amplifier | 8W | 2 W | 1.1 W | 6 W |
|---|---|---|---|---|
| Space Aperture | 7 cm | 10 cm | 13.5 cm | 13.5 cm |
| Ground Aperture | 50 cm | 70 cm | 70 cm | 30 cm |

Due to the high sampling rates required, we have selected pseudo-real-time streaming approach to the data downlink for BHEX. This means that a near-global ground infrastructure is required. Two potential strategies exist: (1) leverage existing astronomical and optical communications telescopes to provide suitable coverage (2) develop the ground stations required. Many ground stations are in development today, with a range of aperture sizes and capabilities [16-20]. By reducing ground station aperture requirements, we allow for a substantially larger number of ground station telescopes to be potential data downlink sites for BHEX. For strategy (2), we calculate that the BHEX data needs can be met by using only 4 ground stations if they can be located in specific locations (e.g., as shown in Figure 4: Hawaii, Greece, Australia, and South America). Additional ground stations (example locations shown in blue) can provide site diversity for cloud cover.

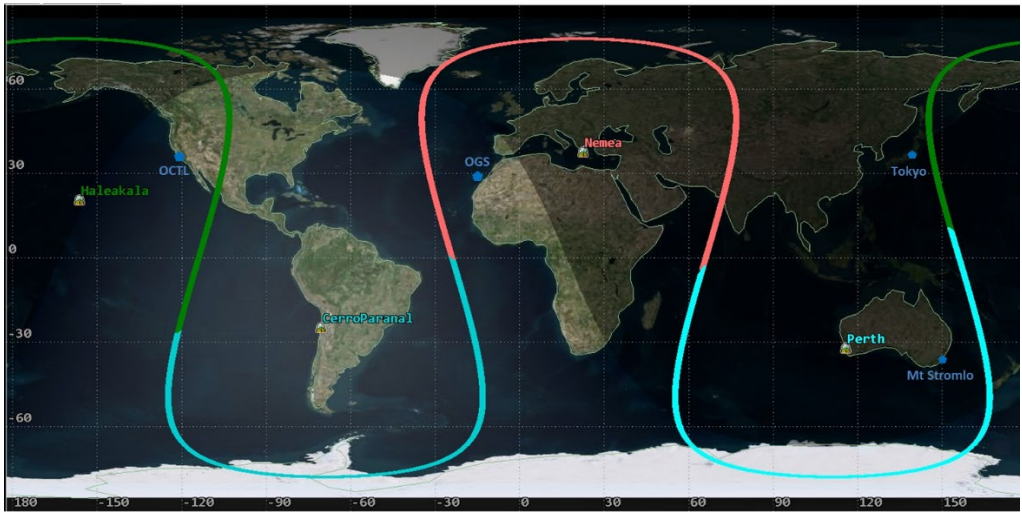

Figure 5. Orbit simulation showing a potential orbit requiring only 4 optical ground stations to support the BHEX science mission. Additional ground stations can be used to improve site diversity.

## 5. CONCLUSION

Over the past several decades, laser communication technologies have been matured through substantial demonstrations covering data rates from <1 Gbps up to 200 Gbps and over a range of orbital regimes spanning LEO through to deep space. Recently, commercial developments have further matured the technology, resulting in a burgeoning laser communication industry. The black hole explorer mission (BHEX) is poised to leverage this recent development to deliver unique insights into black hole physics such as measuring the spin of the black hole and imaging the photon ring for the first time. Here, we present an update to the BHEX lasercom downlink design, leveraging the recently updated requirements to provide a broader trade space allowing us to take advantage of recent and on-going developments in the lasercom community.

## ACKNOWLEDGEMENTS

We would like to acknowledge the BHEX mission proposal team in maturing this system concept. Technical and concept studies for BHEX have been supported by the Smithsonian Astrophysical Observatory and by the University of Arizona. We acknowledge financial support from the Brinson Foundation, the Gordon and Betty Moore Foundation (GBMF-10423), the National Science Foundation (AST-2307887, AST-1935980, and AST-2034306). This project/publication is funded in part by the Gordon and Betty Moore Foundation (Grant #8273.01). It was also made possible through the support of a grant from the John Templeton Foundation (Grant #62286). The opinions expressed in this publication are those of the author(s) and do not necessarily reflect the views of these Foundations. BHEX is funded in part by generous support from Mr. Michael Tuteur and Amy Tuteur, MD. BHEX is supported by initial funding from Fred Ehrsam. In addition, we would like to acknowledge the strong partnership with NASA GSFC which has led to many of the first successful laser communications demonstrations described in this paper.